\documentclass[twocolumn,showpacs,preprintnumbers,amsmath,amssymb,superscriptaddress,floatfix]{revtex4}

\usepackage{graphicx}
\usepackage{dcolumn}
\usepackage{bm}

\begin{document}

\preprint{cond-mat/???????}

\title{Charge Modulations vs. Strain Waves in Resonant X-Ray Scattering}%

\author{P. Abbamonte}
  \affiliation{Department of Physics and Federick Seitz Materials Research Laboratory, University of Illinois, Urbana, IL, 61801}

\date{\today}

\begin{abstract}
A method is described for using resonant x-ray scattering to separately quantify the charge (valence) modulation
and the strain wave associated with a charge density wave.  
The essence of the method is a separation of the atomic form factor into a ``raw" amplitude, $f_R(\omega)$, and a valence-dependent
amplitude, $f_D(\omega)$, which in many cases may be determined independently from absorption measurements.  The advantage of this
separation is that the strain wave follows the average quantity $|f_R(\omega) + <\!\! v \!\!> f_D(\omega)|^2$ whereas the charge modulation
follows only $|f_D(\omega)|^2$.  This allows the two distinct modulations to be quantified separately.  
A scheme for characterizing a given CDW as Peierls-like or Wigner-like follows naturally.
The method is illustrated for an idealized model of a one-dimensional chain.  

\end{abstract}

\maketitle

Charge density waves (CDWs) are pervasive in condensed matter, arising in di- and trichalcogenides, 
conducting polymers, transition metal oxides
etc.  However they may form by a variety of mechanisms.  The best understood is the Peierls mechanism in which a 
gap at the Fermi level opens, and a charge (valence) modulation forms, through the presence of a strain wave.  
The basis for the Peierls mechanism is the electron-phonon interaction which drives pairing between an electron and a hole
with equal but opposite wave vector, ${\bf k}_F$, resulting in a charge modulation with total momentum $2{\bf k}_F$.  
A CDW may also be driven by electron interactions, however, 
an idealized example being the Wigner crystal (WC) ground state of a dilute electron gas, which is driven by direct 
Coulomb\cite{wigner}.  Because of the close analogy between charge density waves and superconductivity \cite{gruner}, 
and the occurance of both in many
systems such as the dichalcogenides and the copper-oxides in which superconductivity is not 
clearly understood, it would be extremely useful to
distinguish between Peierls-like and Wigner-like (i.e. interaction-driven)
mechanisms in practice\cite{WCxtal}.

The main distinction between a Peierls CDW and a WC is the presence of a lattice distortion in the 
former case that is not required in the latter.  
However in any real material even a WC must have at least some incidental lattice distortion if only because of electrostatics.  
Therefore, in practice, distinguishing between these two phenomena is a quantitative rather than a qualitative matter.  

X-ray diffraction is sensitive to both charge modulations and strain waves.  Scattering from both can be enhanced, 
through different resonant mechanisms, by tuning the x-ray photon energy to an atomic core transition, i.e. so-called resonant
x-ray scattering (RXS).  Here a general method is proposed for using RXS to distinguish between scattering from a charge modulation 
and scattering from a strain wave, even when the two coexist.  
This method allows the charge and strain amplitudes to be quantified independently.  
A system classifying a given charge density wave as either Peierls or Wigner in origin
naturally arises from this separation.  The method is demonstrated for the idealized case of a one-dimensional chain.  

\section{Separating the atomic form factor}

In quantum electrodynamics, the cross section for elastic scattering of photons from a material is\cite{sakurai}

\begin{equation}
M_{\bf q} = \frac{e^2}{2mc^2} \, \rho_{\bf q} + \frac{e^2}{\hbar m^2c^2} \sum_n \frac{<0|\hat{{\bf p}} \cdot {\bf A}|n><n|\hat{{\bf p}} \cdot {\bf A}|0>}{\omega-\omega_n+i\gamma}
\end{equation}

\noindent
where $\rho_{\bf q}$ is a Fourier component of the total electron density, {\bf q} is the momentum transfer, $\hat{{\bf p}}$ is the momentum
operator, $\hat{{\bf A}}$ is the photon field operator (vector potential), and $\omega$ is the photon energy.  
The first of these two terms is equivalent to classical Thomson scattering, does not depend on $\omega$, and 
gives rise to the ``normal" dispersion of optical constants in matter, i.e. in SI units 
$\chi_{\bf q} = - r_e \lambda^2 \rho_{\bf q}/4 \pi^2$, where $r_e$ 
is the classical electron radius and $\lambda=2\pi c / \omega$ \cite{henke}.  The second, resonant
term corresponds to scattering via virtual transitions between 
core and valence states, is highly dependent on $\omega$, and is the origin of anomalous dispersion\cite{james}.  
This term, sometimes referred to as the Kramers-Heisenberg formula, 
has been shown to provide direct sensitivity to valence band charge and spin ordering in condensed matter
\cite{gibbs,hannon,kao,castleton,wilkins,dhesi,thomas,schuessler,usScience,usSCO,usLBCO},
provided $\omega$ coincides with an edge threshold.  

How can one distinguish between a charge moduation and a strain wave?  
One might be tempted to just compare the relative sizes of nonresonant 
and resonant scattering on the assumption that a lattice distortion, which translates all the electrons 
in an atom including those in the core, contributes mainly
to the former and valence band effects mainly to the latter.  
Indeed, the Kramers-Heisenburg formula alone is normally used to analyze near-edge scattering data
\cite{castleton,wilkins,thomas,schuessler}.  However one must realize two things.  First, the valence modulation itself
should scatter in the Thomson channel as well since it corresponds to a real, albeit small, charge modulation.  Second,
it was realized long ago\cite{parratt} that the resonant term also contributes to scattering far from the edges
via virtual, off-shell processes.  The optical contants of Cu metal, for example, exhibit no region of `normal' dispersion from 
the infrared up to 100 keV\cite{parratt}.  So the resonant term is omnipresent, a trait that is commonly exploited in  
crystallographic phasing methods\cite{hendrickson}.
More to the point, both the valence and lattice modulations of a CDW scatter in both channels, and a more appropriate 
division of the cross section than Eq. 1 is needed.

Here it is proposed to divide the cross section instead into a `raw' and `valence-dependent' component.  
More specifically, assuming the form factor for an atom, $f$, is 
a function of both $\omega$ and the valence state, $v$, it is proposed to make the separation 

\begin{equation}
f(\omega,v)=f_R(\omega) + v f_D(\omega).
\end{equation}

\noindent
Here $f_R$ is the `raw' part of the atomic scattering factor and $f_D$ describes how it changes
with valence\cite{arbitrary}.  
Like the full $f$, $f_R$ is dimensionless (i.e. `electrons') and converges to $Z^*$ 
as $\omega \rightarrow \infty$\cite{henke}.  $f_R$ contains both 
resonant and nonresonant processes but only those, such as edge jumps, that are independent of the detailed electronic
structure of the atom.

$f_D$ has units of electrons/valence and can be thought of the scattering power of the valence modulation.  
Like $f_R$ this quantity contains both resonant and nonresonant processes, however the nonresonant contribution will be extremely
small.  One can expect $f_D$ to be large only for $\omega$ near the threshold of an absorption edge, where 
the intermediate states involve transitions directly into the valence band.  

There are two advantages to the division in Eq. 2.  First, as will be illustrated in the next section, scattering from a strain wave
will follow only $|f(\omega,<\!\!v\!\!>)|^2$, where $<\!\!v\!\!>$ is the {\it average} atomic valence of the material,
while scattering from the valence modulation follows $|f_D(\omega)|^2$.  By measuring the photon
energy dependence of a CDW reflection, provided $f_D$ and $f_R$ are known, one can determine separately the charge and strain 
amplitudes as well as their relative phase.

Second, and most importantly, in many cases $f_R$ and $f_D$ can be independently determined 
from doping-dependent x-ray absorption (XAS) data.  More specifically, $f$ is related to the absorption coefficient \cite{henke}
by

\begin{equation}
Im[n(\omega,<\!\!v\!\!>)] =-\frac{r_e N \lambda^2}{2 \pi V_c} Im \left [f_R(\omega) + <\!\!v\!\!> f_D(\omega) \right ]
\end{equation}

\noindent
where $n$ is the complex index of refraction, $N$ is the number of atoms in a unit cell, and
$V_c$ is the unit cell volume.  
So, by measuring the absorption coefficient on two or more samples of different $<\!\!v\!\!>$, $Im[f_R(\omega)]$ and
$Im[f_D(\omega)]$ can be determined by solving a system of equations at each value of $\omega$.  The real parts can then by 
determined by Kramers-Kronig transform.  Without having to
appeal to a specific model, then, the charge and strain amplitudes can be determined.  
XAS acts as a reference against which resonant scattering 
measurements can be calibrated.

Use of this procedure rests on several assumptions.  
First, one must assume that 
the scattering processes are local and that an atomic form factor, $f$, is definable.  
This will be valid as long as $v_g \tau \ll \lambda/2$,
where $v_g$ is the group velocity of the core electron-hole pair, $\tau$ is its radiative lifetime, and $\lambda$ is the 
x-ray wavelength.  For measurements near edges with sharp white lines, such as the $L$ edges of the transition metals, 
one can expect this condition to be well satisfied.  
Next one must assume that changes in XAS spectra due to changes in $<\!\! v \!\!>$, determined by comparing
samples with different chemical composition, indeed arise from valence effects and not 
extrinsic phenomena such as changing crystal structures etc.
Next, one must assume that changes in $f$ are small i.e. are linear in $v$, which should
be true as long as the CDW amplitude is not too large.  It is {\it not } necessary that the sample have a rigid or 
noninteracting band structure.
In determining $f_R$ and $f_D$ from XAS one implicitly makes the assumption that the electronic structure of
every point in the CDW corresponds to the {\it average} electronic structure in a sample with that valence.  Finally, one assumes
that x-ray absorption measurements are a good measure of the forward scattering amplitude, i.e. that there are not hidden loss processes
such as photoelectron production.  Violation of this last point has already been observed \cite{usLBCO} but is not a 
significant effect.

\begin{figure}
\includegraphics{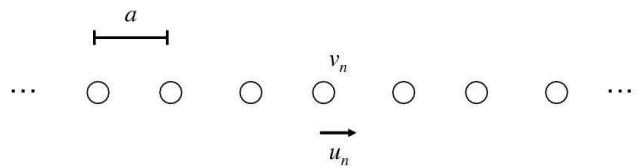}
\caption{
One-dimensional, monatomic chain.  $u_n$ and $v_n$ describe respectively the displacement from equilibrium and valence state of the $n$th 
atom in the lattice. 
}
\end{figure}

\section{One-dimensional chain}

To see how $f_R$ and $f_D$ enter a resonant scattering experiment consider the idealized case of a one-dimensional,
monatomic chain, as shown in Fig. 1.  Imposed upon this chain are a longitudinal strain wave, described by a set of displacements
$u_n$, and a valence modulation, described by a set of valence states, $v_n$.  For simplicity it is assumed that the
modulations are sinusoidal, i.e.

\begin{equation}
u_n = u_0 \cos(k n a)
\end{equation}

\begin{equation}
v_n = \, <\!\! v \!\!> + \, v_0 \cos(k n a + \phi)
\end{equation}

\noindent
where $u_n$ implies that the $n$th atom resides at position $r_n = na + u_n$, where $a$ is the average
lattice parameter.  $u_0$ and $v_0$ are the aplitudes of the
strain and charge waves, respectively, $<\!\! v \!\!>$ is the average
atomic valence, and 
$k$ is the wave vector of the CDW.  The quantity
$\phi$ allows for the fact that, while the two modulations have the same wave vector, there may be a phase difference between
them.  

The integrated intensity of a Bragg reflection is proportional to the square of the scattering amplitude, $\rho_q$,
which is given by

\begin{equation}
\rho_q(\omega)=\frac{1}{N} \sum_{n=1}^N f(\omega,v_n) \; e^{i q u_n}
\end{equation}

\noindent
where $f(\omega,v_n)$ is as defined by eq. (2) and $N = 2\pi/ka$ is the number of sites in the supercell, 
i.e. the dimensionless CDW period \cite{comm}.  $q$ is the momentum transfer along the chain and is restricted 
to discrete values $q=2\pi l/Na$, 
where $l$ is an integer.
If one multiplies out the terms one arrives at four distinct contributions to the scattering amplitude.  

\begin{equation}
\rho_q(\omega)=\rho_q^B(\omega) + \rho_q^v(\omega) + \rho_q^u(\omega) + \rho_q^{uv}(\omega)
\end{equation}

\noindent
The first term,

\begin{equation}
\rho_q^B(\omega) = \frac{1}{N} \; f(\omega,<\!\! v \!\!>) \; \sum_{n=1}^N e^{i q n a},
\end{equation}

\noindent
is the ``Bragg" term and corresponds to resonant x-ray scattering off the average lattice.  
This quantity is independent of $u_0$ and $v_0$ and is nonzero whenever
$q = 2\pi m/a$, where $m$ is an integer, i.e. at the Bragg points of the undistorted chain.
This term always has exactly the value $f(\omega,<\!\! v \!\!>)$, which demonstrates that regular Bragg scattering
has the energy dependence of $|f(\omega,<\!\! v \!\!>)|^2 = |f_R(\omega)+<\!\! v \!\!> f_D(\omega)|^2$, 
i.e. simply tracks the average scattering factor of the atomic lattice.  $\rho^B_k$ is highly resonant but 
nonzero for all values of $\omega$.

The next term 

\begin{equation}
\rho_q^v(\omega) = \frac{1}{N} \, v_0 \, f_D(\omega) \, \sum_{n=1}^N \cos(k n a + \phi) e^{i q n a }
\end{equation}

\noindent
is the ``valence" term and corresponds to resonant x-ray scattering off the valence modulation.  This quantity
is nonzero only for $q=\pm k$ and has the value $\rho_k^v(\omega) = v_0 f_D(\omega) \exp(i\phi)/2$.  This
demonstrates that resonant scattering from the valence modulation tracks only $|f_D(\omega)|^2$ and is proportional to $v_0^2$.  
Since $f_D$ is significant only near an edge, one can expect to see scattering from a valence modulation
only over a narrow energy range near threshold.  This is the experimental signature of a valence modulation in resonant
x-ray scattering.

The third term is the ``strain" term and corresponds to scattering off the lattice distortion.  Provided 
the size of the distortion is small, i.e. $\exp(iqr_n) = \exp(iqna) (1 + iqu_n)$, the strain term has the value

\begin{equation}
\rho_q^u(\omega) = \frac{1}{N} \, i \, q \, u_0 \, f(\omega,<\!\! v \!\!>) \, \sum_{n=1}^N \cos(k n a) e^{i q n a}.
\end{equation}

\noindent
This quantity is also nonzero only if $q=\pm k$ and reduces to 
$\rho_k^u(\omega) = i q u_0 f(\omega,<\!\! v \!\!>)/2$.  Scattering from the strain wave is proportional to
$(k u_0)^2$, as expected, and like the Bragg term follows
$|f(\omega,<\!\! v \!\!>)|^2$ so is visible at all values of $\omega$.  Evidently all structural
scattering is alike in its adherence to the average $f(\omega,<\!\! v \!\!>)$.  
Unlike the Bragg term, however, the strain term occurs at the 
same $q$ as the valence scattering and and the two may coherently interfere.  We will see that, if this interference
is visible, it provides a means to determine the phase shift, $\phi$.

The final term is somewhat unexpected and apparently has not been addressed before.  It is
a mixed term corresponding to coherent displacement of the valence modulation.  Written out, it has the
form

\begin{equation}
\rho_q^{uv}(\omega) = \frac{1}{2N} i q v_0 u_0 f_D(\omega) \sum_{n=1}^N [ \cos(2 k n a + \phi) + cos(\phi) ] e^{i q r_n}
\end{equation}

\noindent
This quantity is nonzero only if $q= \pm 2k$ and, while it arises from both strain and valence 
scattering, it tracks only $f_D$.  $\rho^{uv}$ is not a multiple-scattering effect; it is a sign that if both 
charge and strain modulations are present the total modulation is anharmonic.
Observation of this term, i.e. by tuning
the x-ray energy near threshold and scanning around $2k$, would be a strong validation of our approach.  However this term
is extremely small.

\begin{figure}
\includegraphics{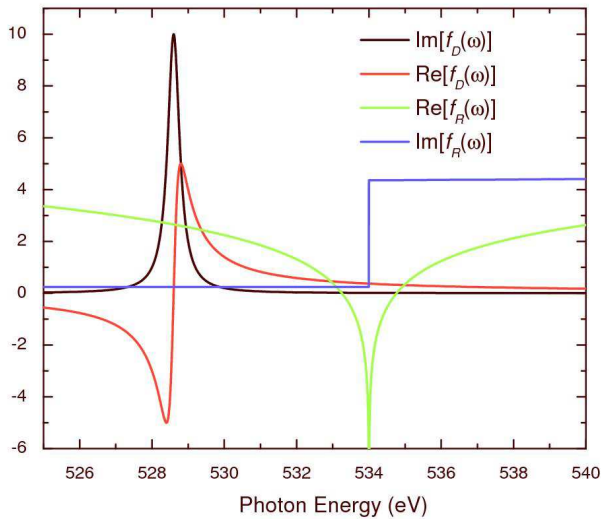}
\caption{
Real and imaginary parts $f_D$ and $f_R$ as determined from the parameter values chosen (see text).  
}
\end{figure}

\section{A generic edge}

The various scattering processes are best illustrated with a specific model of $f_R$ and $f_D$ near an 
absorption edge for a single atom.  
As was argued earlier, $f_R$ describes processes that are material- and valence- independent, such as the
edge jump, and $f_D$ describes only those processes that depend on the atomic valence, $v$.  As an illustrative model of $f_R$, we
consider a generic edge jump, i.e. a dielectric susceptibility whose imaginary part has the form 

\begin{equation}
Im[ \, \chi(\omega) \, ] = \frac{J}{\omega} \; \theta(\omega-\omega_e).
\end{equation}

\noindent
where $\omega_e$ is the edge energy and $J$ is the size of the jump.  
$\chi$ must satisfy the Kramers-Kronig relations so its real part is given by the integral

\begin{equation}
Re[ \, \chi(\omega) \, ] = \frac{2 J}{\pi} \int_{\omega_e}^\infty \frac{d\omega'}{(\omega')^2 - \omega^2} = -\frac{J}{\pi\omega}\log \left | \frac{\omega_e-\omega}{\omega_e+\omega} \right |.
\end{equation}

\noindent
This can be considered a generic analytic model for an idealized edge.
The atomic scattering factor is related to this susceptibility by\cite{batterman}

\begin{equation}
f_R(\omega) = - \frac{V_c \, \omega^2}{r_e \, c^2} \; \chi(\omega)
\end{equation}

\noindent
where $\chi$ is in cgs units.  $V_c$ here is the volume per atom.  Bear in mind that $f$ in general is a tensor,
but for simplicity we will treat it as a scalar here.  
It is important to point out that this quantity actually diverges as $\omega \rightarrow \infty$ but near the 
edge it is well behaved.

Eq. 14 is for an isolated edge.  
However in practice there are other transitions that are far away but still
contribute to $f$.  To account for these we add a constant ``background" scattering factor, $f_0$, i.e.

\begin{equation}
f_R(\omega) \rightarrow f_R(\omega) + f_0
\end{equation}

\noindent
Using the notation $f(\omega) = f^1(\omega)+if^2(\omega)$, writing it out explicitly, $f_R$ has the form

\begin{equation}
f^1_R(\omega) = -\Delta \, \frac{\omega}{\pi \omega_e}\log \left | \frac{\omega_e-\omega}{\omega_e+\omega} \right | + f_0^1
\end{equation}

\begin{equation}
f^2_R(\omega) = \Delta \, \frac{\omega}{\omega_e} \; \theta(\omega-\omega_e) + f_0^2
\end{equation}

\noindent
where $\Delta=V_c J / r_e c^2$.

The valence-dependent form factor, $f_D$, could take on many forms.  In high temperature superconductors, for example,
it actually exhibits
a sign change\cite{usLBCO}.  For present, illustrative purposes, we will simply take it to be a Lorentzian, i.e.

\begin{equation}
f_D(\omega) = \frac{A}{\omega-\omega_0-i\gamma}
\end{equation}

\noindent
In principle $f_D$ should have an energy-independent component representing Thomson scattering from the valence modulation, 
but in situations of interest this is small.

For numerics, we will take the parameter values $f_1^0 = 9.50$, $f_2^0 = 0.246$, $\Delta = 4.114$, $A=2$ eV, and $\gamma=0.2$ eV,
$\omega_e=534$ eV, and $\omega_0=528.6$ eV. While it is intended that this illustration be general these parameters are quite a 
good model of the oxygen K edge.  $f_R$ and $f_D$ for these parameter values are plotted in Fig. 2.

For the CDW itself we use the parameters $v_0 = 0.1$ electron, $u_0 = 0.1 a$, $k=2\pi/4a$, and $<\!\! v \!\!> = 0.12$.  
The scattered intensity at $q=k$ is given by square of the total scattering amplitude, $|\rho^u_k(\omega)+\rho^v_k(\omega)|^2$.
The strain and charge scattering amplitudes can coherently interfere, so must be added before squaring.  The resulting quantity 
depends explicitly on the phase, $\phi$.  

The quantity $|\rho^u_k(\omega)+\rho^v_k(\omega)|^2$ is plotted against energy in Fig. 3.  For this plot $\phi$ is taken to be zero.
Spectra are shown for a pure charge wave,
a pure strain wave, and a composite wave.  Notice that scattering from the charge-only wave is nonzero only in the region in which $f_D$, displayed in
Fig. 2, is nonzero.  
The strain-only wave, in contrast, is visible at (almost) all energies.  
This affirms ones intuition that scattering from structural distortions should be visible at all energies, but the valence modulation
only near the edge.
If both modulations are present the lineshape is not simply
the sum of the two because of the non-trivial dependence on the phase factor, $\phi$.

To illustrate this phase dependence the line shape is plotted in Fig. 4 for vaious values of $\phi$.  
Notice that not only the intensity of the various features but in fact the entire spectral lineshape depends sensitively 
on $\phi$.  Therefore, if both strain and charge scattering are simultaneously visible, and $f_R$ and $f_D$ are determined 
independently from XAS, it should be possible to objectively determine the phase from a one-parameter fit to this shape.

\begin{figure}
\includegraphics{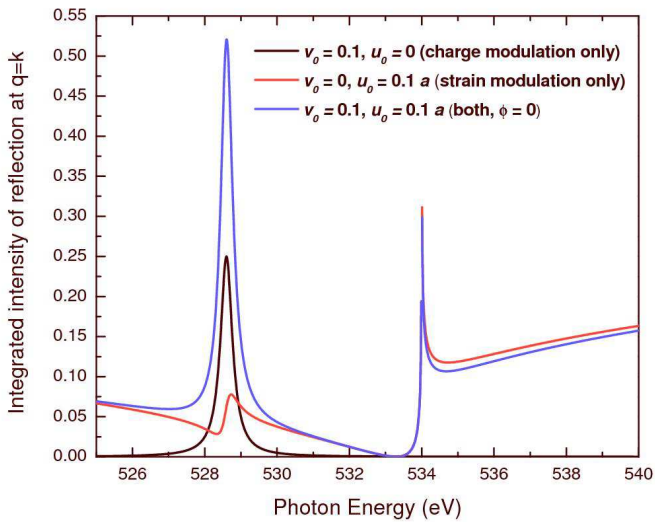}
\caption{
Plots of the integrated intensity of resonant x-ray scattering at $q=k$, i.e. $|\rho^u_k(\omega)+\rho^v_k(\omega)|^2$.  Here $\phi=0$.
(black line) Charge modulation only.  (red line) Strain wave only.  (blue line) Both charge and lattice modulations present.
Charge scattering is localized near
threshold but the strain wave, while resonant, is visible at all energies.  Notice that the lattice distortion has a 
strong intensity max at the edge jump due to the cusp in the real part of $f_R$.  
}
\end{figure}

\begin{figure}
\includegraphics{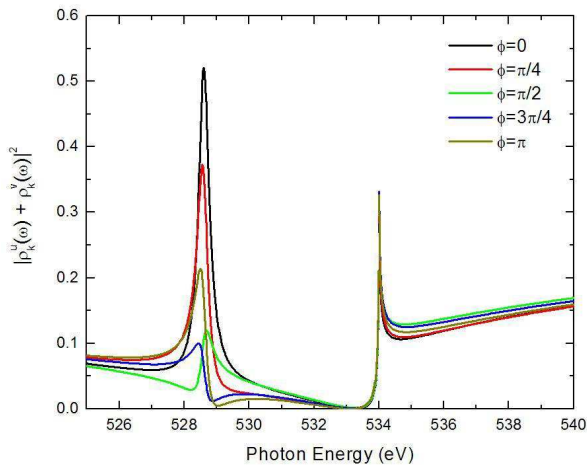}
\caption{
Plots of the integrated intensity of resonant x-ray scattering at $q=k$, i.e. $|\rho^u_k(\omega)+\rho^v_k(\omega)|^2$ for 
various values of the phase between charge and strain modulations, $\phi$.  Notice that the spectral shape changes with 
$\phi$.  If $f_D$ and $f_R$ are known from absorption data, $\phi$ can be determined through analysis of this shape.
}
\end{figure}

\section{peierls vs. wigner cdws}

Distinguishing between a Peierls CDW and a 
more exotic CDW driven by many-body interactions, such as a stripe phase or a Wigner crystal, is a quantitative rather
than a qualitative matter.  This is because in any real material even a CDW driven purely by electron-electron interactions, because
of electrostatics, must still be accompanied by a lattice distortion, though it may be small.  
Identifying a given CDW as either Peierls-like or Wigner-like 
requires a quantitative comparison between the charge and lattice amplitudes.  Once
$u_0$ and $v_0$ have been determined by the procedure just outlined, then, a quantity of great interest is the ratio

\begin{equation}
W = \frac{v_0}{u_0},
\end{equation}

\noindent
This quantity has units of 
length$^{-1}$ and describes the degree to which the CDW is Wigner-like, i.e. diven by many-body interactions rather than a
Peierls distortion.  For example, a perfect
Wigner crystal with no lattice distortion at all would have $W=\infty$.  A typical Peierls CDW on the other hand, 
such as that in NbSe$_3$, has a lattice
distortion of approximately $u_0 \sim 0.01 \AA$ and a charge amplitude of $v_0 \sim 0.1$, giving $W=10 \AA^{-1}$.  
We propose that a CDW with $W$ less than about 20 should be considered a Peierls CDW.  
If $W > 100$ the CDW probably arises at least partly from many-body effects and should be considered ``exotic".  
It is likely that CDWs can exist over the entire continuum of values of $W$.  It would be particularly enlightening
to determine the $W$ values for several CDW materials, such as the copper-oxides and the 
dichalcogenides, that also exhibit superconductivity.

\section{summary}

A descripton of resonant x-ray scattering was introduced in which the atomic scattering 
factor is divided into raw- and valence-dependent amplitudes, $f_R$ and $f_D$.
The advantage of this division is that resonant x-ray scattering from 
the strain wave component of a CDW tracks the average form factor 
$|f_R(\omega) + <\!\! v \!\!> f_D(\omega)|^2$ whereas the charge (valence)
scattering tracks only $|f_D(\omega)|^2$.  In many cases $f_R$ and $f_D$ 
can be independently determined from x-ray absorption
measurements on materials with different average valence, combined with Kramers-Kronig analysis. This provides a means to 
separately quantify the charge and strain components of a CDW.  In this framework one can define a quantity ``$W$" which provides
a quantitative means to characterize a given CDW as Peierls-like, Wigner-like, or anywhere on the continuum between.  

The author thanks M. V. Klein for a critical reading of the manuscript.
This work was funded by the Materials Sciences and Engineering Division,
Office of Basic Energy Sciences, U.S. Department of Energy under grant No.
DE-FG02-06ER46285.  

\bibliography{oChainPRB-v2}

\begin{thebibliography}{22}
\bibitem[{WCx()}]{WCxtal}
\bibinfo{note}{In this manuscript the term `Wigner crystal' (WC) will refer to
  any charge density wave that is driven pricipally by many-body interactions
  rather than a Peierls distortion, a density wave in the dilute electron gas
  driven by direct Coulomb being just one example.}
\bibitem[{arb()}]{arbitrary}
\bibinfo{note}{$v$ is defined with respect to a reference valence state of the
  atom. This reference is arbitrary, since the transformtion $v \rightarrow v +
  \Delta v$ may be accomodated by the redefinition $f_R \rightarrow f_R +
  \Delta v f_D$.}
\bibitem[{com()}]{comm}
\bibinfo{note}{For the sake of illustration it is assumed that $N$ is an
  integer, i.e. the CDW is commensurate. However the method also works for an
  incommensurate CDW.}
\end{thebibliography}

\end{document}